 \documentclass[12pt]{article}
 \usepackage{graphicx}
 \usepackage{amsfonts,epsf}

 \catcode`\@=11 \@addtoreset{equation}{section}\catcode`\@=12

 \def\barr{\left(\begin{array}}
 \def\earr{\end{array}\right)}
 \def\nqq{\hspace{-2em}}
 
 \def\ber#1{\begin{eqnarray}\label{#1} \nqq}
 \def\eer{\end{eqnarray}}
 \newcommand{\beq}[1]{\begin{equation}\label{#1}}
 \newcommand{\eeq}{\end{equation}}
 \newcommand{\bear}[1]{\begin{eqnarray}\label{#1}}
 \newcommand{\ear}{\end{eqnarray}}
 \newcommand{\be}{\begin{equation}}
 \newcommand{\ee}{\end{equation}}
 \newcommand{\ba}{\begin{eqnarray}}
 \newcommand{\ea}{\end{eqnarray}}
 
 \newcommand{\fnm}{\footnotemark}
 \newcommand{\fnt}{\footnotetext}
 \newcommand{\eps}{\varepsilon}
 \newcommand{\tri}{\Delta}
 \newcommand{\sign}{ \mbox{\rm sign} }
 \newcommand{\e}{ \mbox{\rm e} }
 \newcommand{\N}{ {\mathbb N} }
 \newcommand{\R}{ {\mathbb R} }
 \newcommand{\nn}{\nonumber}

\begin{document}
 \thispagestyle{empty}

 \vskip 35mm

 \begin{center}
 {\large\bf  On billiard approach in multidimensional cosmological models}

  \vspace{15pt}

 \normalsize\bf
 V.D. Ivashchuk\fnm[1]\fnt[1]{ivashchuk@mail.ru}
 and  V.N. Melnikov\fnm[2]\fnt[2]{melnikov@phys.msu.su}

 \vspace{5pt}

 \it Center for Gravitation and Fundamental Metrology,
 VNIIMS, 46 Ozyornaya ul., Moscow 119361, Russia  \\

 Institute of Gravitation and Cosmology,
 Peoples' Friendship University of Russia,
 6 Miklukho-Maklaya ul.,  Moscow 117198, Russia \\

 \end{center}
 \vskip 10mm

\begin{abstract}

 A short overview of the billiard approach for cosmological-type
 models with $n$ Einstein factor-spaces
  is presented. We start with the billiard representation
 for  pseudo-Euclidean  Toda-like systems of cosmological
 origin. Then we consider  cosmological model with
 multicomponent ``perfect-fluid'' and
 cosmological-type model with composite branes.
 The second one describes  cosmological
 and spherically-symmetric configurations
 in a theory with scalar fields and fields of forms.
 The conditions for appearance of
 asymptotical Kasner-like and oscillating behaviors
 in the limit $\tau \to +0$ and $\tau \to + \infty$
 (where $\tau$ is a ``synchronous-type'' variable) are
 formulated (e.g. in terms of inequalities on Kasner
 parameters). Examples of billiards related
 to the hyperbolic Kac-Moody algebras
 $E_{10}$, $AE_3$ and $A_{1,II}$ are
 given.

 \end{abstract}

 \vskip 10mm

 PACS numbers:  04.50.-h.  \\

 \vskip 30mm

 \pagebreak

\section{Introduction}
\setcounter{equation}{0}

This paper is devoted to  the billiard approach for
multidimensional cosmological-type  models defined on the manifold
 $(u_{-},u_{+}) \times M_{1} \times \ldots \times M_{n}$, where
 all $M_{i}$ are Einstein spaces. This approach was inspired by
Chitre's idea \cite{Chit} of explanation the
 BLK-oscillations \cite{BLK} in the mixmaster model (based on
 Bianchi-IX metric) \cite{Mis0,Mis1} by using a simple triangle billiard
 in the Lobachevsky space $H^2$.

Let us briefly overview the ``history'' of the billiard approach
in multidimensional cosmology. In multidimensional case the
billiard  representation for cosmological model with
 multicomponent  ``perfect'' fluid  was  introduced in
 \cite{IKM1,IKM2,IMb0}. In  our paper \cite{IMb0} the finiteness of
the billiard volume  was formulated in terms of the so-called
illumination problem,  the inequalities on Kasner parameters were
 written  and the ``quantum billiard'' was also considered.
 The mathematical quintessence of the derivation
 of billiard representation was considered in our
 paper \cite{IMRC} devoted to
 pseudo-Euclidean Toda-like systems of cosmological origin.

  The  billiard approach for multidimensional models
  with scalar fields and fields of forms (e.g. for supergravitational
  ones) was suggested in  our paper \cite{IMb1}. This
   paper contains the inequalities on Kasner parameters that
  played a key role in the proof of Damour and  Henneaux conjecture
    on ``chaotic'' behavior in superstring-inspired
  (e.g.  supergravitational)  models \cite{DamH1} (for  more
  detailed explanation see  review article \cite{DHN}).
  We note that at the moment   there are no examples of
  cosmological ($S$-brane) configurations in $D = 11$
  supergravity with diagonal (or block-diagonal) metrics that have an oscillating
  behaviour near the singularity (see \cite{IMS}).

  There was also an important observation made
  in \cite{DamH3}: for certain superstring-inspired
  (e.g.  supergravitational)   models the  parts of billiards
  are related to Weyl chambers of certain
  hyperbolic Kac-Moody (KM)  Lie algebras \cite{Kac,Nik,HPS,Sac}.
  This observation   drastically simplifies the proof of the finiteness of the billiard
  volume. Using this approach the old well-known result
  of Demaret, Henneaux and  Spindel \cite{DHSp}   on critical dimension
  of pure gravity  was explained using hyperbolic algebras in \cite{DamHJN}.

  It should be noted   that earlier few examples
  of hyperbolic KM algebras   were considered
   in our papers (with co-authors)  \cite{IMBl,GrI,IKM} in
  a context of exact solutions  with branes, see also
    \cite{IMsigma-08}.

  Here we overview some our results from \cite{IMb0,IMb1,IMRC}
  with a certain  generalization, e.g. we consider two asymptotical
  regions when i) $\tau \to +0$  (labelled  by $\eps = + 1$)
  and ii) $\tau \to + \infty$ ($\eps = - 1$)
 where $\tau$ is a ``synchronous-type'' variable (it may be time
 variable or radial variable for spherically-symmetric configurations
 in the model with form fields and scalar fields) and give some
 concrete examples, related to hyperbolic KM algebras.

\section{Billiard representation for pseudo-Euclidean
        Toda-like systems of cosmological origin}
        \setcounter{equation}{0}

Here we consider a pseudo-Euclidean Toda-like system described by
the following Lagrangian
 \begin{equation} \label{1.1}
 L = {L}(z^{a}, \dot{z}^{a}, {\cal N}) = \frac{1}{2} {\cal N}^{-1}
 \eta_{ab} \dot{z}^{a} \dot{z}^{b} -  {\cal N} {V}(z),
 \end{equation}
where ${\cal N} >0 $ is the Lagrange multiplier (modified lapse function),
 \\
 $(\eta_{ab})=  {\rm diag}(-1,+1, \ldots ,+1)$  is the matrix of minisuperspace
metric, $a,b = 0, \ldots , N-1$, and
 \begin{equation} \label{1.2}
 {V}(z) = \sum_{\alpha=1}^{M} A_{\alpha} \exp(u^{\alpha}_a z^a)
 \end{equation}
 is the potential,  all $A_{\alpha} \neq 0$.

 We consider the behavior of the dynamical system (\ref{1.1}) for
 $N \geq 3$ in the limit
 \begin{equation} \label{1.3}
 z^2  \equiv  -(z^0)^2 + (\vec{z})^2   \rightarrow  -\infty, \qquad
 z =(z^0, \vec{z}) \in {\cal V}_{-\eps}, \end{equation}

where ${\cal V}_{-\eps} \equiv \{(z^0, \vec{z}) \in \R^N : \eps
        z^0  < -  |\vec{z}| \}$
 is the lower or upper light cone for $\eps =
 +1, -1$ respectively. The limit (\ref{1.3}) implies
 $z^0 \rightarrow  \mp \infty$ for $\eps = \pm 1$.
  For $\eps = + 1$ it describes (under certain additional
 assumptions imposed)  approaching  to the singularity in
 corresponding cosmological models.

We impose the following restrictions on vectors $u^{\alpha} =
 (u^{\alpha}_0, \vec{u}^{\alpha})$ in the potential (\ref{1.2})

 \begin{eqnarray} \label{1.5}
 && 1) \ A_{\alpha} > 0 \ {\rm if} \ (u^{\alpha})^2 =
 -(u^{\alpha}_0)^2 + (\vec{u}^{\alpha})^2 > 0;
 \\ \label{1.6}
 && 2) \ \eps u^{\alpha}_0 > 0  \ {\rm if} \ (u^{\alpha})^2 \leq 0.
 \end{eqnarray}

Let us consider the behavior of the dynamical system, described by
the Lagrangian (\ref{1.1}) for $N \geq 3$ in the limit
 (\ref{1.3}). We restrict the Lagrange system (\ref{1.1}) on ${\cal
  V}_{- \eps}$.

  {\bf Remark 1.} In a general case the shifted cone
   ${\cal V}_{-\eps}(\eta) \equiv \{(z^0, \vec{z}) \in \R^N : \eps
  (z^0 - \eta^0)  < -  |\vec{z} - \vec{\eta}| \}$
   should be considered, where
  $\eta = (\eta^0, \vec{\eta})$. Here we put $\eta = 0$ for
  simplicity.

 We introduce  an analogue of the Misner-Chitre coordinates in
 $\cal{V}_{- \eps}$:
  \begin{eqnarray} \label{2.2}
  &&z^0 = - \eps \exp(- \eps y^0) \frac{1 + \vec{y}^2}{1 - \vec{y}^2},
  \\ \label{2.3}
  &&\vec{z} = - 2 \eps \exp(- \eps y^0) \frac{ \vec{y}}{1 - \vec{y}^2},
  \end{eqnarray}
  $|\vec{y}| < 1$, and fix the  gauge
 \be
 \label{2.9}
  {\cal N} =   \exp(- 2 \eps y^0) = - z^2.
 \ee

 In what follows we consider  $(N-1)$-dimensional Lobachevsky space $H^{N-1}$
 realized as a unit ball
   $H^{N-1} = D^{N-1} \equiv \{ \vec{y}= (y^1, \ldots, y^{N-1}): |\vec{y}| < 1
                             \}$
  with the metric $h = 4 \delta_{ij} (1 - \vec{y}^2)^{-2} dy^i \otimes dy^j $.

 The set of indices
 $\Delta_{+}  \equiv \{ \alpha : (u^{\alpha})^2 > 0 \}$
  defines a billiard $B$ in $H^{N-1}$:
  \be \label{2.22}
  B = \bigcap_{\alpha \in
 \Delta_{+}} {B}(u^{\alpha}), \ee
  where the subset
  ${B}(u^{\alpha})$ consists of points
  $\vec{y} \in D^{n-1}$ obeying:

  i)  $(\vec{y} - \vec{v}^{\alpha})^2 >
 (\vec{v}^{\alpha})^2 - 1$ for $\eps u^{\alpha}_{0} > 0$;

  ii)  $(\vec{y} - \vec{v}^{\alpha})^2 <
 (\vec{v}^{\alpha})^2 - 1$ for $\eps u^{\alpha}_{0} < 0$;

  iii)  $\eps \vec{y} \vec{u}^{\alpha} > 0$
        for $ u^{\alpha}_{0} = 0$;

       $\alpha \in \Delta_{+}$.

 Here
  \be \label{2.24}
 \vec{v}^{\alpha} = -
 \vec{u}^{\alpha}/u^{\alpha}_{0} \ {\rm for} \ u^{\alpha}_{0} \neq 0.
 \ee

 $B$ is an open domain. Its boundary
 $\partial B = \bar{B} \setminus B$ is formed by certain parts of
 $m_{+} = |\Delta_{+}|$  $(N-2)$-dimensional planes or spheres with  centers in
the points (\ref{2.24})  ($|\vec{v^{\alpha}}| > 1$) and radii
  $r_{\alpha} = \sqrt{(\vec{v}^{\alpha})^2 - 1}$.

  It may be  shown  that in
 the limit  $y^0 \rightarrow - \eps \infty$ (or, equivalently, in the
 limit (\ref{1.3})) the Lagrange equations for the Lagrangian (\ref{1.1}) with the
 gauge fixing (\ref{2.9}) (under restrictions (\ref{1.5}) and
 (\ref{1.6}) imposed)   are reduced
   to  Lagrange equations for  the Lagrangian
 \be \label{2.29}
  L_{B} =  \frac{1}{2} {h_{ij}}(\vec{y})
 \dot{y}^{i} \dot{y}^{j} -  {V}(\vec{y},B), \ee

 where
 \ba \label{2.21}
  {V}(\vec{y},B) \equiv &0, &\vec{y} \in B,
  \nonumber \\
  &+ \infty, &\vec{y}
  \in D^{N-1} \setminus B,
 \ea
 is a potential describing $m_{+}$ billiard walls.
 The  $y^0$-variable is separated: $y^0 = \omega (t - t_0)$,
 ($\omega \neq 0$ , $t_0$  are constants) and
 the energy constraint  $E_{B}  = \omega^2/2$ should be imposed
 (for $\eps = 1$ see \cite{IMb0,IMRC}).

We put $\omega > 0$, then the limit $t \rightarrow - \eps \infty$
corresponds to (\ref{1.3}).  When $\Delta_{+} = \emptyset$, we
have $B = D^{N-1}$ and  Lagrangian (\ref{2.29}) describes  the
geodesic  flow  on the Lobachevsky space $H^{N-1}$. In this case
there are two families of non-trivial geodesic solutions  (i.e.
 ${y}(t) \neq const$): lines or semi-circles  orthogonal to the
boundary $S^{N-2}$ \cite{IMb0}.

 When  $\Delta_{+} \neq \emptyset$ Lagrangian
 (\ref{2.29}) describes a motion of a particle  of  unit  mass,  moving
 in the billiard $B$.  For cosmological models (see
 next section) the geodesic motion in $B$
 corresponds to a ``Kasner epoch'' and the reflection from the
 boundary corresponds to a change of Kasner epoch.

When billiard $B$ has the infinite volume there are  open zones
at  the infinite  sphere $|\vec{y}| =1$. After a finite number of
reflections from the boundary the  particle (in a general case)
moves toward  one  of  these  open zones. For the corresponding
cosmological model we get the ``Kasner-like'' asymptotical
behavior in the limit  $t \rightarrow - \eps \infty$ .

 For (non-empty) billiards with finite volume
  the motion of the particle describes
  an ``oscillatory-like'' asymptotical
  behaviour of the corresponding cosmological model
  in the limit  $t \rightarrow - \eps \infty$.

In \cite{IMb0}  we proposed a simple ``illumination'' criterion
for the finiteness of the volume of $B$, which in the extended
form is:

 {\bf Proposition 1}. {\em The billiard $B$ (\ref{2.22}) has a
 finite volume if and only if:
 the point-like sources of light located at
 the points $\vec{v}^{\alpha}$ (\ref{2.24}) for $\eps u^{\alpha}_0
  > 0$, the sources at infinity
  $- \infty \eps \vec{u}^{\alpha} $ for $u^{\alpha}_0 = 0$ and
 ``anti-sources'' located at points  (\ref{2.24}) for $\eps u^{\alpha}_0 < 0$
 illuminate the unit sphere $S^{N-2}$.
 For ``anti-source'' the shadowed domain
 coincides with the illuminated domain for the usual source located
 at the same point (and vice versa)}.

This proposition was proved in \cite{IMb0} for usual sources of
light, when all $u^{\alpha}_0 > 0$ and $\eps = +1$.

 The problem of illumination of a convex body in a vector space by
 point-like sources for the first time was considered in
 \cite{32,33}. For the case of $S^{N-2}$ this problem is equivalent
 to the problem of covering the spheres with spheres
 \cite{34,35}. There exists a topological bound on the number of
 usual point-like sources $m_{+}$ illuminating sphere $S^{N-2}$
 \cite{33}:
 \be \label{2.36}
  m_{+} \geq N.
 \ee

\section{Billiard representation for a cosmological model with
         $m$-component perfect-fluid}

In this section we consider a cosmological model describing the
evolution of $n$ Einstein spaces in the presence of $m$-component
perfect-fluid matter. The metric of the model
 \begin{equation} \label{f1.1.1}
  g=-\exp[2{\gamma}(t)]dt \otimes dt +
  \sum_{i=1}^{n} \exp[2{x^{i}}(t)] \hat{g}^{i},
 \end{equation}
 is defined on the  manifold
 \begin{equation}  \label{f1.1.2}
  M = (t_{-},t_{+}) \times M_{1} \times \ldots \times M_{n},
 \end{equation}
where the manifold $M_{i}$ with the metric $g^i$ is an Einstein
space of dimension $d_i$,
 ${R_{m_{i}n_{i}}}[g^i] = \xi_{i} g^i_{m_{i}n_{i}}$,
  $i = 1, \ldots ,n $; $n \geq 2$.
Here and in what follows $\hat{g}^{i} = p_{i}^{*} g^{i}$ is the
pullback of the metric $g^{i}$  to the manifold  $M$ by the
canonical projection: $p_{i} : M \rightarrow  M_{i}$,
 $i = 1,\ldots, n$.

The energy-momentum tensor is adopted in the following form
 \begin{eqnarray}  \label{f1.1.4}
  && T^{M}_{N} = \sum_{\alpha =1}^{m} T^{M (\alpha)}_{N}, \\ \label{f1.1.5}
  &&(T^{M (\alpha)}_{N})= {\rm diag}(-{\rho^{(\alpha)}}(t),
  {p_{1}^{(\alpha)}}(t) \delta^{m_{1}}_{k_{1}},
  \ldots , {p^{(\alpha)}_{n}}(t) \delta^{m_{n}}_{k_{n}}),
  \end{eqnarray}
 $\alpha =1, \ldots ,m$, with the conservation law constraints imposed:

 \begin{equation}    \label{f1.1.6}
 \bigtriangledown_{M} T^{M (\alpha)}_{N}=0 \end{equation}
 $\alpha=1, \ldots ,m-1$.

 The Einstein equations
 \begin{equation}   \label{f1.1.7}
 R^{M}_{N}-\frac{1}{2}\delta^{M}_{N}R=\kappa^{2}T^{M}_{N}
 \end{equation}
 ($\kappa^{2}$ is the multidimensional gravitational constant) imply
 $\bigtriangledown_{M} T^{M}_{N}=0$ and consequently
 $\bigtriangledown_{M} T^{M (m)}_{N}=0$.

 We suppose that for any $\alpha$-th component of matter
 pressures in all spaces are proportional to a density
  \begin{equation}    \label{f1.1.8}
  p_i^{(\alpha)}(t) = \left(1- \frac{u_i^{(\alpha)}}{d_i} \right)\rho^{(\alpha)}(t),
  \end{equation}
 where $u_i^{(\alpha)}$ are constants,
  $i=1, \ldots,n$;  $\alpha=1, \ldots , m$.

The conservation law constraint (\ref{f1.1.6})  reads
 $\dot{\rho}^{(\alpha)}+
 \sum_{i=1}^{n}d_i\dot{x}^{i}(\rho^{(\alpha)} + p_{i}^{(\alpha)})=0$
 and hence using  (\ref{f1.1.8}) we get
  \begin{equation}  \label{f1.2.9}
   {\rho^{(\alpha)}}=  A^{(\alpha)}
   \exp[-2d_i x^i + u_i^{(\alpha)}x^i],
 \end{equation}
 where $A^{(\alpha)}$ are constant numbers.

It was shown in \cite{IM5,IMb0} that the Einstein equations
(\ref{f1.1.7}) for the metric (\ref{f1.1.1}) and  the
energy-momentum tensor (\ref{f1.1.4}), (\ref{f1.1.5}) with
(\ref{f1.1.8}) are equivalent to the Lagrange equations for the
following degenerate Lagrangian

 \begin{equation}  \label{f1.2.18}
 L = \frac{1}{2} \exp(- \gamma + {\gamma_{0}}(x)) G_{ij}
 \dot{x}^{i}\dot{x}^{j}- \exp( \gamma
 - {\gamma_{0}}(x)) V(x),
 \end{equation}

 where
 \begin{equation}  \label{f1.2.3}
  \gamma_{0} \equiv \sum_{i=1}^{n} d_{i}x^{i},
 \end{equation}
 and
 \begin{equation}  \label{f1.g}
 G_{ij} = d_i \delta_{ij}- d_i d_j
  \end{equation}
 are  components of  minisuperspace metric \cite{IMZ},  and

 \begin{equation} \label{b1.2.16}
  V = {V}(x) = -\frac{1}{2}\sum_{i=1}^{n} \xi_{i} d_{i}
  \exp(-2x^{i}+2 {\gamma_{0}}(x)) + \sum_{\alpha=1}^{m} \kappa^{2}
  A^{(\alpha)} \exp(u_i^{(\alpha)}x^i)
 \end{equation}

 is  the potential.

The relation (\ref{b1.2.16}) may be also presented in the form
 \begin{equation} \label{b1.2.17}
  V = \sum_{\alpha= 1}^{\bar{M}}  A_{\alpha} \exp(u_i^{(\alpha)}x^i),
 \end{equation}
 where  $\bar{M} = m + n$; $A_{\alpha} = \kappa^{2} A^{(\alpha)}$,
    $\alpha= 1, \ldots , m$; $A_{m+i} =- \frac{1}{2} \xi_{i} d_{i}$
  and
  \begin{equation} \label{b1.2.18}
   u_j^{(m+i)}= 2(- \delta^i_j + d_j),
  \end{equation}
   $i,j=1,\ldots,n$.

  {\bf Diagonalization.}

  The minisuperspace metric   $G = G_{ij}dx^{i} \otimes dx^{i}$
 has the pseudo-Euclidean signature $(-,+, \ldots ,+)$ \cite{IMZ}, i.e. there exists
 a linear transformation
 \begin{equation} \label{f1.3.2}
  z^{a}=e^{a}_{i}x^{i},
 \end{equation}
 diagonalizing the minisuperpace metric:
  $G= \eta_{ab} dz^{a} \otimes dz^{b}$
 where

 $(\eta_{ab})=(\eta^{ab}) \equiv {\rm diag}(-1,+1, \ldots ,+1)$,
  $a,b = 0, \ldots ,n-1$.

 Like in \cite{IMZ} we put
 \begin{equation} \label{b1.2.30}
  z^0 = e^{0}_{i} x^i = q^{-1} d_i x^i, \qquad q = [(D-1)/(D-2)]^{1/2}.
 \end{equation}
 For the volume scale factor
   $v = \exp(\sum_{i=1}^{n} d_{i}x^{i}) = \exp(qz^0)$
 we get $v \to +0$ for $z^0 \to - \infty$ and $v \to +\infty$ for
  $z^0 \to + \infty$.

 Let us  denote
  \begin{equation} \label{b1.2.35}
   u^{\alpha}_a  = e_{a}^{i} u^{(\alpha)}_i,
  \end{equation}
 $a = 0, \ldots , n-1$, where $(e_{a}^{i}) = (e^{a}_{i})^{-1}$.

  Then the Lagrangian (\ref{f1.2.18}) written in $z$-variables
   is coinciding with   the Lagrangian (\ref{1.1}) with $N = n$,
   $M \leq n +m $
   and ${\cal N}=\exp( -\gamma_0 + \gamma)>0$.

 It was shown in \cite{IMb0} that
  \begin{equation} \label{b1.2.36}
  u^{\alpha}_0  =  \left( \sum_{i=1}^{n} u^{(\alpha)}_i \right) / q(D-2).
  \end{equation}
 and
 \begin{equation} \label{b1.2.38}
 (u^{\alpha})^2 = (u^{(\alpha)}, u^{(\alpha)}) ,
 \end{equation}

where $(u, v) = G^{ij}u_iv_j$,
 \begin{equation} \label{b1.2.g}
  G^{ij} = \frac{\delta^{ij}}{d_{i}}+ \frac{1}{2-D}
 \end{equation}
  are components of the matrix inverse to $(G_{ij})$,
  $D = 1 + \sum_{i=1}^{n} d_i$ is the dimension of the manifold
 (\ref{f1.1.2}.)

 For the curvature $u$-vectors we get

 \begin{equation} \label{b1.2.37}
   \qquad u^{m + j}_0 = 2/q > 0,
 \end{equation}
  and
 \begin{equation} \label{b1.2.39}
  (u^{m + j})^2 =
  4 \left( \frac{1}{d_j} - 1 \right) < 0,
 \end{equation}
 for $d_j > 1$, $j= 1,\ldots,n$. For $d_j = 1$ we have $\xi_{j}
 = A_{m+j} = 0$.

  {\bf Billiard restrictions.}

  The restrictions 1) and 2)
  (see (\ref{1.5}) and (\ref{1.6})) read (due to
  (\ref{b1.2.36})-(\ref{b1.2.39})):

   \begin{eqnarray} \label{b1.3.3}
  && 1) \ A^{(\alpha)} > 0  \  {\rm if} \ (u^{(\alpha)}, u^{(\alpha)}) >  0;  \\
  \label{b1.3.4}
  && 2a) \ \eps \sum_{i=1}^{n} u^{(\alpha)}_i  > 0 \
         {\rm if} \ (u^{(\alpha)}, u^{(\alpha)}) \leq  0; \\
  \label{b1.3.5}
    &&  2b) \ \xi_{j} =  0, \ j =1, \dots, n, \ {\rm if}  \ \eps = -1.
  \ea

  The last condition means that  all factor spaces $M_i$ should be Ricci-flat
  when the case $\eps = -1$ is studied.

  {\bf Kasner-like parametrization.}

 Let the billiard has an infinite volume and hence there are
 open zones at infinity.
 It may be shown (along a line as it was done in  \cite{IMb0} when all
 $u^{\alpha}_0 > 0$ and $\eps = +1$)
 that the geodesic motion in $H^{n-1}$ towards one of these
  zones corresponds to
 ``Kasner-like''  asymptotical behaviour of the metric
(\ref{f1.1.1}) in the limit when  $\tau  \rightarrow + 0$ for
$\eps = +1$ or $\tau  \rightarrow + \infty$ for $\eps = -1$.
 Here $\tau$ is the synchronous time variable.
 The asymptotical form of the metric reads
 \ba \label{b1.4.33}
 &&g_{as} = - d\tau \otimes d\tau +
 \sum_{i=1}^{n} A_i \tau^{2 \alpha^i}
 \hat{g}^i,  \\
 \label{b1.4.34}
 && \sum_{i=1}^{n} d_i \alpha^i = \sum_{i=1}^{n} d_i (\alpha^i)^2 = 1,
 \ea
where $A_i > 0$ are constants.  Here the Kasner parameters obey
the following inequalities:

 \be \label{b1.4.37}
 \sum_{i=1}^{n} \eps  u_i^{(\nu)} \alpha^i   > 0,
 \ee
  $\nu \in \Delta_{+}$.

 The  Kasner set $\alpha = (\alpha^i)$ is in one-to-one correspondence with
 the unit vector $\vec{n} \in S^{n-2}$:
  $\alpha^i =  e^i_a n^a / q$,  $(n^a) = (1, \vec{n})$.

The criterion of the finiteness of the billiard volume (see
Proposition 1) may be reformulated in terms of inequalities on the
Kasner-like parameters.

{\bf Proposition 2.} {\em The (non-empty) billiard $B$
(\ref{2.22}) has a finite volume if and only if the set of
relations (\ref{b1.4.34}), (\ref{b1.4.37}) is inconsistent}.

This proposition may be proved  along a line as it was done in
 \cite{IMb0} when all $u^{\alpha}_0 > 0$ and $\eps = +1$. For
finite (non-zero) billiard volume  we get a never ending
asymptotical oscillating behaviour.

{\bf Remark 2.} Let all factor spaces are Ricci-flat, i.e.
$\xi_{j} =  0, \ j =1, \dots, n$. It is not difficult to verify
that for a fixed diagonalization procedure
(\ref{f1.3.2})-(\ref{b1.2.35}) the billiard $B$ (\ref{2.22}) is
unchanged when the following transformation of parameters is
performed:

 \be \label{1.r2} u^{(\alpha)}_i \mapsto - u^{(\alpha)}_i, \qquad
 \eps \mapsto - \eps, \ee

 $i=1, \ldots,n$, $\alpha=1, \ldots , m$.
 In terms of the parameters
 $w_i^{(\alpha)} = 1- \frac{u_i^{(\alpha)}}{d_i}$
 (i.e. $p_i^{(\alpha)} =
   w_i^{(\alpha)} \rho^{(\alpha)}$), the first formula in relation (\ref{1.r2})
 reads:
   $w^{(\alpha)}_i \mapsto \hat{w}^{(\alpha)}_i =
    2 -  w^{(\alpha)}_i$.

 Thus,  for a given billiard $B$ describing
the behaviour near the singularity (either Kasner-like or
never-ending oscillating one) as $\tau \to + 0$ ($\eps = +1$) we
get the same billiard $B$ for $\tau \to + \infty$ ($\eps = -1$)
when the $u$-parameters   are replaced according to (\ref{1.r2}).

{\bf Remark 3.} For a fixed diagonalization procedure
(\ref{f1.3.2})-(\ref{b1.2.35}) the billiard $B$ from Section 2 is
unchanged when the following transformation of parameters is done:

   \be \label{1.r3}
   u^{(\alpha)}_i \mapsto  \lambda_{(\alpha)}
   u^{(\alpha)}_i,
   \ee

  where $\lambda_{(\alpha)} > 0$,
   $i=1, \ldots,n$,  $\alpha=1, \ldots , m$. Here $\eps$
   is unchanged.

 In terms of the $w$-parameters the relation (\ref{1.r3})
 reads
 $w^{(\alpha)}_i \mapsto \hat{w}^{(\alpha)}_i
 = \lambda_{(\alpha)} (w^{(\alpha)}_i - 1) + 1$.

 {\bf Collision formula.}

 Let the billiard $B$ has a finite volume. In this case
 we get a never ending oscillation behaviour
 in the asymptotical regime. In a period between two collisions
 with potential walls we have a Kasner-like
 relations for the metric (\ref{b1.4.33})
  with $\alpha$-parameters obeying (\ref{b1.4.34}).
 It may be shown (along the line as it was done in \cite{Ierice}
 for $S$-brane solutions)   that the set  of Kasner parameters
 $(\alpha^{'i})$ after the collision with the $s$-th wall
  (corresponding to the $s$-th component), $s \in \Delta_{+}$,
  is defined by the Kasner set before the collision
 $(\alpha^{i})$ according to the following formula
  \beq{gcl-a}
   \alpha^{'i} =
               \frac{\alpha^i - 2 u^{(s)}(\alpha) u^{(s)i}(u^{(s)},u^{(s)})^{-1}}
               {1 - 2 u^{(s)}(\alpha)
               (u^{(s)},u^{\Lambda})(u^{(s)},u^{(s)})^{-1}},
  \eeq
   $i = 1, \ldots, n$.    Here  $u^{(s)} (\alpha) = u_i^{(s)} \alpha^i $,
   $u^{(s)i} = G^{ij} u^{(s)}_j$ and $u^{\Lambda}_i = 2 d_i$.

 \section{Billiard representation for a cosmological-type  model with branes }

 Now, we consider the model governed by the action
 \bear{i2.1}
 S =&&
  \int_{M} d^{D}z \sqrt{|g|} \{ {R}[g] - 2 \Lambda - h_{\alpha\beta}
 g^{MN} \partial_{M} \varphi^\alpha \partial_{N} \varphi^\beta
 \\ \nn
 && - \sum_{a \in \Delta}
 \frac{\theta_a}{n_a!} \exp[ 2 \lambda_{a} (\varphi) ] (F^a)^2_g \},
 \ear

where $g = g_{MN} dz^{M} \otimes dz^{N}$ is a metric on the
manifold $M$, ${\dim M} = D$, $\varphi=(\varphi^\alpha)\in \R^l$
is a vector from dilatonic scalar fields,
 $(h_{\alpha\beta})$ is a positive-definite symmetric
 $l\times l$ matrix ($l\in \N$),  $\theta_a  \neq 0$,
 $ F^a =  dA^a =\frac{1}{n_a!} F^a_{M_1 \ldots M_{n_a}}
 dz^{M_1} \wedge \ldots \wedge dz^{M_{n_a}} $
is a $n_a$-form ($n_a \geq 2$) on a $D$-dimensional manifold $M$,
 $\Lambda$ is a cosmological constant
and $\lambda_{a}$ is a $1$-form on $\R^l$ :
 $\lambda_{a} (\varphi) =\lambda_{a \alpha}\varphi^\alpha$,
 $a \in \Delta$, $\alpha=1,\ldots,l$.
In (\ref{i2.1}) we denote
  $|g| = |\det (g_{MN})|$, $(F^a)^2_g =
  F^a_{M_1 \ldots M_{n_a}} F^a_{N_1 \ldots N_{n_a}} g^{M_1 N_1}
  \ldots g^{M_{n_a} N_{n_a}}, $ $a \in \Delta$, where $\Delta$ is
 some finite set.
 In  models with one time all $\theta_a
  =  1$  when the signature of the metric is $(-1,+1, \ldots, +1)$.

We consider the manifold
 \beq{i2.10}
 M = (u_{-}, u_{+})  \times M_{1} \times \ldots \times M_{n},
 \eeq
with the metric
 \beq{i2.11}
 g= w e^{2 \gamma(u)} du \otimes du  +
     \sum_{i=1}^{n} e^{2 x^i(u)} \hat{g}^i,
 \eeq
where $w = \pm 1$,
 $g^i  = g^{i}_{m_{i} n_{i}}(y_i) dy_i^{m_{i}} \otimes
 dy_i^{n_{i}}$ is an Einstein metric on $M_{i}$  satisfying
 $R_{m_{i}n_{i}}[g^i ] = \xi_{i} g^i_{m_{i}n_{i}}$,
 $m_{i},n_{i}=1, \ldots, d_{i}$; $\xi_{i}= {\rm const}$,
 $i=1,\ldots,n$.
The functions $\gamma, x^{i} : (u_{-}, u_{+}) \rightarrow \R $ are
smooth. We denote $d_{i} = {\rm dim} M_{i}$; $i = 1, \ldots, n$;
 $D = 1 + \sum_{i = 1}^{n} d_{i}$. Here $u$ is a variable
 by convention called ``time''.

 We consider any manifold $M_{i}$ to
 be oriented and connected. Then the volume $d_i$-form
 \beq{i2.14}
  \tau_i  \equiv \sqrt{|g^i(y_i)|}
  \ dy_i^{1} \wedge \ldots \wedge dy_i^{d_i},
 \eeq and signature parameter

 \beq{2.15}
 \eps(i)  \equiv {\rm sign}( \det (g^i_{m_i n_i})) = \pm 1
 \eeq

 are correctly defined for all $i=1,\ldots,n$.

 Let $\Omega = \Omega(n)$  be a set of all non-empty subsets of
 $\{  1, \ldots,n \}$. The number of elements in $\Omega$ is $|\Omega| =
 2^n - 1$. For any $I = \{ i_1, \ldots, i_k \} \in \Omega$, $i_1 <
 \ldots < i_k$, we denote

 \bear{i2.16}
 \tau(I) \equiv \hat{\tau}_{i_1}  \wedge \ldots \wedge \hat{\tau}_{i_k},  \\
 \label{i2.17}
 \eps(I) \equiv \eps(i_1) \ldots \eps(i_k),  \\
  \label{i2.19}
  d(I) \equiv  \sum_{i \in I} d_i.
  \ear

Here $\hat{\tau}_{i} = p_{i}^{*} \hat{\tau}_{i}$ is the pullback
of the form $\tau_i$  to the manifold  $M$ by the canonical
projection: $p_{i} : M \rightarrow  M_{i}$, $i = 1,\ldots, n$. We
also put $\tau(\emptyset)= \eps(\emptyset)= 1$ and
$d(\emptyset)=0$.

For fields of forms we consider the following composite
electromagnetic ansatz
\ber{2.1.1}
 F^a=\sum_{I\in\Omega_{a,e}}{\cal F}^{(a,e,I)}+
 \sum_{J\in\Omega_{a,m}}{\cal F}^{(a,m,J)}
\eer
 where
  \bear{2.1.2}
 {\cal F}^{(a,e,I)}=d\Phi^{(a,e,I)}\wedge\tau(I), \\
 \label{2.1.3} {\cal F}^{(a,m,J)}=
 e^{-2\lambda_a(\varphi)}*(d\Phi^{(a,m,J)} \wedge\tau(J)) \ear

 are elementary forms of electric and magnetic types respectively,
 $a\in\tri$, $I\in\Omega_{a,e}$, $J\in\Omega_{a,m}$ and
 $\Omega_{a,v} \subset \Omega$, $v = e,m$. In (\ref{2.1.3})
 $*=*[g]$ is the Hodge operator on $(M,g)$.

 For scalar functions we put
 \ber{2.1.5}
 \varphi^\alpha=\varphi^\alpha(u), \quad
 \Phi^s=\Phi^s(u),
 \eer
$s\in S$.

Here and below
 \ber{2.1.6} S=S_e \sqcup S_m, \quad
  S_v=\sqcup_{a\in\tri}\{a\}\times\{v\}\times\Omega_{a,v}, \eer
  $v=e,m$
  ($\sqcup$ means the union of non-intersecting sets). The
  set $S$ consists of elements $s=(a_s,v_s,I_s)$, where $a_s \in
 \tri$ is the colour index, $v_s = e, m$ is the electro-magnetic index and
 set $I_s \in \Omega_{a_s,v_s}$ describes the location of a brane.

Due to (\ref{2.1.2}) and (\ref{2.1.3})
  $d(I)=n_a-1, \quad d(J)=D-n_a-1$,
  for  $I \in \Omega_{a,e}$ and $J \in \Omega_{a,m}$
 (i.e. in electric and magnetic cases, respectively).

{\bf Restrictions on brane intersections.} Here we put two
restrictions on sets of branes that guarantee the block-diagonal
form of the energy-momentum tensor and the existence of the
sigma-model representation (without additional constraints):

 \beq{2.2.2a}
  {\bf (R1)} \quad d(I \cap J) \leq
  d(I) - 2, \eeq
  for any $I,J \in\Omega_{a,v}$, $a\in\tri$, $v= e,m$
 (here $d(I) = d(J)$)
  and
 \beq{2.2.3a}
 {\bf (R2)} \quad d(I \cap J)  \neq 0,
 \eeq
 for any $I\in\Omega_{a,e}$, $J \in \Omega_{a,m}$, $a\in\tri$.

 It follows from \cite{IMC} that equations of motion for the model
 (\ref{i2.1}) and the Bianchi identities:  $d{\cal F}^s=0$,
  $s \in S_m$, for fields from (\ref{i2.11}),
 (\ref{2.1.1})-(\ref{2.1.5}), when Restrictions (\ref{2.2.2a}) and (\ref{2.2.3a})
 are  imposed, are equivalent to equations of motion for the
  $1$-dimensional $\sigma$-model with the action

 \beq{4.1.1}
 S_{\sigma} =  \int du {\cal  N}^{-1}
 \biggl\{ \hat{G}_{AB}\dot \sigma^A \dot \sigma^B
 + \sum_{s\in  S}\eps_s \exp[-2U^s(\sigma)](\dot\Phi^s)^2 -2{\cal
  N}^{2}V_{w}\biggr\}, \eeq

where $\dot X \equiv dX/du$,
 \beq{4.1.2}
 V_{w} = -w\Lambda\e^{2\gamma_0(x)}+ \frac{w}{2}
 \sum_{i =1}^{n} \xi_i d_i \e^{-2 x^i + 2 {\gamma_0}(x)}
 \eeq

is the potential,
 $\gamma_0(x) \equiv \sum_{i=1}^nd_i x^i$ and
 ${\cal N}=\exp( -\gamma_0 + \gamma)>0$.

  In (\ref{4.1.2})  $(\sigma^A)=(x^i,\varphi^\alpha)$, the index set $S$ is
 defined in (\ref{2.1.6}),
 \beq{4.G}
 (\hat{G}_{AB}) = {\rm diag}(G_{ij}, h_{\alpha \beta})
 \eeq
 is the matrix of a minisuperspace metric  and
  \beq{2.2.11}
   U^s(\sigma) =
   U_A^s \sigma^A = \sum_{i \in I_s} d_i x^i - \chi_s \lambda_{a_s}(\varphi), \quad
  (U_A^s) =  (d_i \delta^{i}_{I_s}, -\chi_s \lambda_{a_s \alpha})
  \eeq
  are the so-called $U$-(co)vectors, $s=(a_s,v_s,I_s)$.
 Here $\chi_e=+1$ and $\chi_m=-1$;
 \ber{2.2.12}
 \delta^i_{I}=\sum_{j \in I}\delta^i_{j}
 \eer
 is an indicator of $i$ belonging to $I$: $\delta^i_{I}=1$ for
 $i\in I$ and $\delta_{iI}=0$ otherwise; and

 \beq{2.2.13a}
 \eps_s=  \eps(I_s) \theta_{a_s} \ {\rm for} \ v_s = e; \qquad
 \eps_s = -\eps[g] \eps(I_s) \theta_{a_s} \ {\rm for} \ v_s = m,
 \eeq

 $s\in S$, and $\eps[g] \equiv \sign\det(g_{MN})$.

 Now we integrate the Lagrange equations corresponding to $\Phi^s$
(i.e. the ''Maxwell equations'' for $s\in S_e$ and Bianchi
identities for $s\in S_m$):

 \bear{3.2.14}
 \frac
  d{du}\left( {\cal  N}^{-1} \exp(-2U^s(\sigma))\dot\Phi^s\right)=0
 \Longleftrightarrow \dot\Phi^s = Q_s {\cal  N} \exp(2U^s(\sigma)),
 \ear

where $Q_s$ are constants, $s \in S$.  We put $Q_s \ne 0$ for all
$s \in S$.  For fixed $Q=(Q_s,s \in S)$ the Euler-Lagrange
equations for the action (\ref{4.1.1})  corresponding to
 $(\sigma^A)=(x^i,\varphi^{\alpha})$, when equations
(\ref{3.2.14}) are substituted, are equivalent to Lagrange
equations for the Lagrangian

 \beq{3.2.16}
  L = {\cal  N}^{-1} \hat G_{AB}\dot \sigma^A \dot \sigma^B - {\cal  N} V,
 \eeq
where
 \beq{3.2.17}
    V = V_{w} + \frac12  \sum_{s\in S}  \eps_s Q_s^2
    \exp[2U^s(\sigma)]
 \eeq
 and the matrix $(\hat G_{AB})$ is defined in
(\ref{4.G}).
 This potential may be rewritten as
\beq{3.2.17a}
    V =   \sum_{r \in S_{*}} A_r   \exp[2U^r(\sigma)],
 \eeq
 where $S_{*} = S \sqcup \{ \Lambda \}  \sqcup  \{1, \dots, n \} $,
 $A_s = \frac{1}{2} \eps_s Q_s^2$, $s \in S$,
 $A_{\Lambda} = - w \Lambda$ and $A_i =  \frac{w}{2} \xi_i d_i$,
 $i = 1, \dots, n$. Here $U^{\Lambda}_i =  d_i$,
  $U^{i}_j = - \delta^i_j + d_j$ and all other components are zero.

 We remind that \cite{IMC}
 \beq{b2.39o}
 (U^{\Lambda},U^{\Lambda}) = - q^2  < 0, \qquad
 (U^{j},U^{j}) = \frac{1}{d_j} - 1 < 0,
 \end{equation}
 for $d_j < 1$ and
 \beq{4.3}
  (U^s,U^{s})= d(I_s ) \left(1 +\frac{d(I_s)}{2-D} \right)
  + \lambda_{a_s \alpha} \lambda_{a_s \beta} h^{\alpha \beta}.
  \eeq

 Here and in what follows
  \beq{4.3a}
 (U,U') = \hat{G}^{AB}U_A U'_{B},
  \eeq
  where the matrix $(\hat G^{AB}) = {\rm diag} (G^{ij}, h^{\alpha \beta})$ is
 inverse to $(\hat G_{AB})$ and $(h^{\alpha \beta}) = (h_{\alpha
  \beta})^{-1}$.

{\bf Diagonalization.}

  The minisuperspace metric   $\hat{G} = \hat{G}_{AB}d \sigma^{A} \otimes d \sigma^{B}$
 has the pseudo-Euclidean signature $(-,+, \ldots ,+)$
 (since $(h_{\alpha\beta})$ is a positive-definite), i.e. there exists
 a linear transformation
 \begin{equation} \label{b1.3.2}
  z^{a}=e^{a}_{A} \sigma^{A},
 \end{equation}
  diagonalizing the minisuperpace metric:
  $\hat{G} = \eta_{ab} dz^{a} \otimes dz^{b}$
  where \\
  $(\eta_{ab})=(\eta^{ab}) \equiv diag(-1,+1, \ldots ,+1)$,
  $a,b = 0, \ldots , N-1$; $N = n+ l$.

 Like in the previous section we put
 \begin{equation} \label{b2.2.30}
  z^0 = e^{0}_{A} \sigma^A = q^{-1} d_i x^i.
 \end{equation}

 Let us  denote
  \begin{equation} \label{b2.2.35}
  \hat{U}^{r}_a  = e_{a}^{A} U^{r}_A
  \end{equation}
 $a = 0, \ldots , N - 1$, where $(e_{a}^{A}) = (e^{a}_{A})^{-1}$.
 We get
  \begin{equation}
  \label{b2.2.35a}
  \eta^{ab} \hat{U}^{r}_a \hat{U}^{r'}_b = (U^{r}, U^{r'})
   \end{equation}
  for all $r,r'$.

  Then, the Lagrangian (\ref{3.2.16}) written in $z$-variables
   is coinciding (after a suitable redefinitions of parameters
  e.g. $2\hat{U}^{r}_a  = u^{r}_a$, and indices) with
   Lagrangian (\ref{1.1}) where $N = n + l$ and $M \leq |S| + n +
  1$.

 In what follows we will use the following inequalities
 \cite{IMb1}
  \beq{b2.2.37o}
  \hat{U}^{\Lambda}_0 = q > 0 , \qquad \hat{U}^{j}_0 = 1/q > 0, \eeq
  $j= 1,\ldots,n$, and
   \beq{b2.2.38o}
   \hat{U}^{s}_0 = d(I_s)/\sqrt{(D-2)(D-1)}  > 0,
  \eeq
  $s \in S$.

 {\bf Billiard restrictions.}

 For  $U$-vectors the restrictions 1) and 2)
  (see (\ref{1.5}), (\ref{1.6})),  imply
  (due to (\ref{b2.39o}) and  (\ref{b2.2.35a})-(\ref{b2.2.38o}) )
  \bear{4.1}
    1)  \qquad \qquad \eps_s > 0 \ {\rm for} \ (U^s, U^s) > 0; \\
  \label{4.2}
   2a)  \qquad  {\rm all} \ (U^s, U^s) > 0 \ {\rm if}  \ \eps = -1; \\
  \label{4.3b}
   2b) \quad \qquad \quad \xi^{j}  = \Lambda = 0 \ {\rm if} \ \eps
            =  -1;
  \ear
   $s \in S$, $j= 1,\ldots,n$.

  {\bf Remark 4.} For $\theta_a =
  1$, $a \in \Delta$, and $\eps[g] = -1$, the
  inequality $\eps_s > 0$ means that all $\eps(I_s) = 1$,
   i.e. a brane  with positive $(U^s,U^s)$ should have
  either Euclidean worldvolume or that containing even number of ``times''.
  The inequality $(U^s, U^s) > 0$ is satisfied in a special case
  when  $d(I_s) < D - 2$.

  Let $S_{+} = \{s \in S: (U^s,U^s) > 0 \}$.
 In this model the branes with $s \in S_{+}$ are the only matter
 components responsible for asymptotical formation
 of billiard walls (when $\eps = -1$ we should put $S =  S_{+}$ ).

 {\bf Kasner-like solutions and oscillating behaviour.}

 Let a billiard $B$ has an infinite volume  and hence there are
 open zones at infinity.

  Then we get  ``Kasner-like''  asymptotical behaviour of the metric and scalar
 fields  in the limit when  $\tau  \rightarrow + 0$ (for
 $\eps = +1$) or $\tau  \rightarrow + \infty$ (for $\eps = -1$):
 \bear{4.8}
 &&g = w d\tau \otimes d\tau +
                        \sum_{i=1}^{n} A_i \tau^{2 \alpha^i} \hat{g}^i,
 \\   \label{4.9}
 &&\varphi^{\beta} =  \alpha^{\beta} \ln \tau + \varphi^{\beta}_0,
 \\   \label{4.10}
 && \sum_{i=1}^{n} d_i \alpha^i = \sum_{i=1}^{n} d_i (\alpha^i)^2 +
 \alpha^{\beta} \alpha^{\gamma} h_{\beta \gamma}= 1,
\ear

 where $w = \pm 1$, $A_i > 0$,  $\varphi^{\beta}_0$ are constants
  $i = 1, \ldots, n$; $\beta, \gamma = 1, \ldots, l$.  The
 the  set of Kasner parameters $\alpha = (\alpha^{A}) =
  (\alpha^{i}, \alpha^{\gamma})$ obeys the relations

 \beq{4.12}
 \eps U^s(\alpha) = \eps U_A^{s} \alpha^A =
 \eps \left( \sum_{i\in  I_s} d_i\alpha^i  -
 \chi_s\lambda_{a_s \gamma}\alpha^{\gamma} \right) > 0,
 \eeq
  $s \in S_{+}$.
 Thus, we get  $U^s(\alpha)>0$ for
  $\tau  \rightarrow + 0$ and
  $U^s(\alpha) < 0$ for $\tau  \rightarrow + \infty$, $s \in S_{+}$.

 Here $\tau$ is the ``synchronous time''  variable. The set
 of Kasner parameters $\alpha$ is in one-to-one correspondence with
 the unit vector $\vec{n} \in S^{N-2}$:
 $\alpha^A =  e^A_a n^a / q$, where $(n^a) = (1, \vec{n})$.

 {\bf Proposition 3.} {\em The (non-empty) billiard $B$ (\ref{2.22}) has a finite
  volume if and only if there are no $\alpha$  satisfying the
  relations  (\ref{4.10}) and (\ref{4.12}).}

 This proposition may be proved just along the line as it was done in
 \cite{IMb1} for the case $\eps = +1$ when all $d(I_s) < D - 2$ and
all $\eps_s = +1$. We remind that for finite (non-zero) billiard
volume we get a never ending asymptotical oscillating behaviour.

{\bf Collision formula and scattering law.}

 It was shown in \cite{Ierice} that the set
 of Kasner parameters
 $(\alpha^{'A})$ after the collision with the $s$-th wall
  is defined by the Kasner set before the collision
 $(\alpha^{A})$ according to the following formula
  \beq{gcl}
   \alpha^{'A} =
               \frac{\alpha^A - 2 U^s(\alpha) U^{sA}(U^s,U^s)^{-1}}
               {1 - 2 U^s(\alpha) (U^s,U^{\Lambda})(U^s,U^s)^{-1}}.
  \eeq
   Here  $U^s$ is a brane  co-vector corresponding to the $s$-th wall and
   $U^{sA} = \hat{G}^{AB} U^{s}_B $.
   In the special case of one  scalar field and  $1$-dimensional factor-spaces
   (i.e.  $l= d_i =1$) this formula was suggested earlier
  in \cite{DamH1}. Another special case of the collision law
  for multidimensional multi-scalar cosmological model
  with exponential potentials  was considered in \cite{DIMbil}.

    Recently in \cite{IMsl-08}
    the exact $S$-brane solution (either electric or magnetic)
    in a model with $l$ scalar fields and one antisymmetric form
    of rank $m \geq 2$   was considered. All factor spaces
     $M_1, ..., M_n$ were supposed to be Ricci-flat and $\Lambda =
     0$.  A special solution  governed
    by the function $cosh$ was singled out. It was shown that this special solution
    has  Kasner-like asymptotics in the  limits $\tau \to  + 0$ and
    $\tau \to  + \infty$,  where $\tau$ is the synchronous time variable.
    A relation between two sets of Kasner parameters
    $\alpha_{\infty}$ and  $\alpha_{0}$ was found. Remarkably, this
    relation,  named as ``scattering law'' formula,
     coincided with the ``collision law''  formula (\ref{gcl}).

    \section{Examples of billiards related to hyperbolic Kac-Moody algebras}

  The special class of billiards with finite volumes   occurs
  in the model (\ref{i2.1}) when
  \beq{5.C}
       2 \frac{(U^s,U^{s'})}{(U^{s'},U^{s'})}=  A_{s s'},
  \eeq
   $s, s' \in S_{+}$, where   $A = (A_{s s'})$ is the (generalized) Cartan matrix
   for hyperbolic Lorentzian Kac-Moody (KM) algebra ${\cal G}$, see \cite{Kac,HPS}.
   In this case the   billiard is a projection of the Weyl chamber on the Lobachevsky
   space $H^{N-1}$ ($|S_{+}| = N$ is the rank of ${\cal G}$). Here
   \cite{IMC}
     \beq{5.prod}
  (U^s,U^{s'})=
  d(I_s \cap I_{s'})+\frac{d(I_s)d(I_{s'})}{2-D} +\chi_s \chi_{s'}
  \lambda_{a_s \alpha}\lambda_{a_{s'} \beta} h^{\alpha \beta}.
  \eeq

    We remind that hyperbolic KM algebras are by definition Lorentzian Kac-Moody
    algebras with the property that removing any node from their Dynkin
    diagram leaves one with a Dynkin diagram of the affine or finite type.
    For exact solutions with branes corresponding
    non-singular KM algebras (e.g. hyperbolic ones) see \cite{IMsigma-08}.

   Sometimes the billiard $B$ may be cut into several identical (isomorphic)
   parts $B_k$ so that any $B_k$ corresponds to the hyperbolic algebra ${\cal G}$.
   Then the volume of $B$ is finite.

  {\bf Example 1.} Let us consider model (\ref{i2.1}) with $D =
  11$, $l=0$ (scalar fields are absent) and $F^I$
  are 4-forms ($a=I$), $I \in \Delta = \{I \in \{1, \dots, 10 \}: |I| =3 \}$.
  Thus, $\Delta$ contains all subsets of $\{1, \dots, 10 \}$ having 3
  elements. The number of such forms is $120$. We consider
  the non-composite electric $S$-brane ansatz when all $d_i = 1$.
  In this case the billiard $B$ for $\eps  = + 1$ belongs to
  Lobachevsky space $H^9$. It may be
  cut on several identical parts $B_k$ (using the so-called symmetry
  walls) such that any $B_k$ corresponds to the hyperbolic algebra $E_{10}$
  with the Dynkin diagram pictured on Fig. 1.
  The volume of $B$ is finite. This billiard  appeared
  for composite electric $S$-brane configuration with non-diagonal metric in $D = 11$
  supergravity \cite{DamH1,DamH3}.

 \begin{center}
\begin{picture}(90,20)
\put(5,5){\line(1,0){80}} \put(5,5){\circle*{1}}
\put(15,5){\circle*{1}} \put(25,5){\circle*{1}}
\put(35,5){\circle*{1}} \put(45,5){\circle*{1}}
\put(55,5){\circle*{1}} \put(65,5){\circle*{1}}
\put(75,5){\circle*{1}} \put(85,5){\circle*{1}}
\put(65,5){\line(0,1){10}} \put(65,15){\circle*{1}}
\put(5,2){\makebox(0,0)[lc]{1}} \put(15,2){\makebox(0,0)[lc]{2}}
\put(25,2){\makebox(0,0)[lc]{3}} \put(35,2){\makebox(0,0)[lc]{4}}
\put(45,2){\makebox(0,0)[lc]{5}} \put(55,2){\makebox(0,0)[lc]{6}}
 \put(65,2){\makebox(0,0)[lc]{7}} \put(75,2){\makebox(0,0)[lc]{8}}
 \put(85,2){\makebox(0,0)[lc]{9}}
 \put(68,15){\makebox(0,0)[lc]{10}}
 \end{picture} \\[5pt]
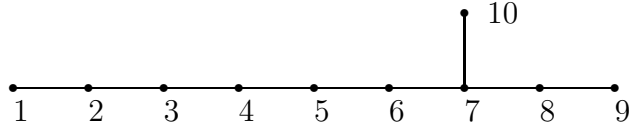
 \small Fig. 1. \it Dynkin diagram for $E_{10}$ hyperbolic KM
  algebra
 \end{center}

  For the model with multicomponent perfect fluid an analogous relation

   \beq{5.Cu}
       2 \frac{(u^{(s)},u^{(s')})}{(u^{(s')},u^{(s')})}=  A_{s s'},
  \eeq
   $s, s' \in \Delta_{+}$, gives us an example of billiard with finite volume
   corresponding to the hyperbolic KM algebra ${\cal G}$ (with
     the Cartan matrix $A = (A_{s s'}))$.

   {\bf Example 2.}
    The billiard with a finite volume corresponding to the hyperbolic algebra $E_{10}$
    occurs for $D = 11$ in the cosmological model with ten-component perfect fluid
    and 1-dimensional factor spaces ($d_i = 1$)  when  the fluid $u$-vectors
    are the following:    $u^{(j)}_i = \lambda \eps (\delta^{j}_i -
    \delta^{j +1}_i)$ for $j = 1, \dots, 9 $, and $u^{(10)}_i = \lambda \eps ( \delta^{8}_i
    + \delta^{9}_i + \delta^{10}_i)$, $i = 1, \dots, 10$,
    where $\lambda > 0$ and $\eps = \pm 1$.
    This $10$-component anisotropic fluid model with equations
    of state parametrized by $\lambda > 0$ and $\eps = \pm 1$
    leads to the oscillating    behaviour of scale factors
    for  $\tau \to +0$ if  $\eps = + 1$
    and for $\tau \to + \infty$ if $\eps = - 1$.

   {\bf Example 3.} The billiard (with a finite volume)   corresponding to the
   hyperbolic KM algebra (that is number 7 in classification of
   \cite{Sac} and $A_{1,II}$ in classification
   of \cite{Nik})) with the Cartan matrix
   \beq{B.1}
  \left(A_{ss'}\right)=\left(\begin{array}{ccc}
   2 & -2 & -2 \\
  -2 &  2 & -2 \\
  -2 & -2 &  2
  \end{array}\right)
  \eeq
    occurs in $D = 4$ for the cosmological model with 3-component perfect fluid
    and 1-dimensional factor spaces ($d_i = 1$)  when  the fluid $u$-vectors
    are the following:    $u^{(j)}_i = 2 \lambda \eps \delta^{j}_i$,
    for $i,j = 1,2,3$,  where $\lambda > 0$ and $\eps = \pm 1$.
    This billiard is coinciding with the Chitre's billiard
    for Bianchi-IX model ($\eps = +1, \lambda = 2$).
    See Fig. 4 in \cite{IMRC}.

   {\bf Example 4.} Another example of a billiard (with a finite volume)   corresponding to the
   hyperbolic KM algebra $AE_3 = {\cal F}_3$  ($A_{1,0}$ in classification
   of \cite{Nik}) with the Cartan matrix
   \beq{A.1}
  \left(A_{ss'}\right)=\left(\begin{array}{ccc}
   2 & -1 &  0 \\
  -1 &  2 & -2 \\
   0 & -2 &  2
  \end{array}\right)
  \eeq
    occurs in the 4-dimensional cosmological model with 3-component perfect fluid
    and 1-dimensional factor spaces ($d_i = 1$)  when  the fluid $u$-vectors
    have the following form:    $u^{(j)}_i =  \lambda \eps (\delta^{j}_i - \delta^{j+1}_i)$ ,
    for $j = 1,2$, and $u^{(3)}_i = 2 \lambda \eps  \delta^{3}_i$,
    $i = 1,2,3$,  where $\lambda > 0$ and $\eps = \pm 1$. This
    billiard may be obtained from that for Bianchi-IX model by cutting it into six identical
    parts (using three lines crossing the center).

 \section{Conclusions}

 Here we  reviewed  the billiard approach for cosmological-type
 models with $n$ Einstein factor-spaces.
 First, we have considered a derivation of the
 billiard approach  for  pseudo-Euclidean  Toda-like systems
 of cosmological  origin.  Then we have applied the billiard scheme
 to the cosmological model with  multicomponent ``perfect-fluid'' and to
 cosmological-type model with composite branes.   We have also formulated
 the conditions for appearance  of asymptotical Kasner-like behaviour and
 ``never ending'' oscillating behavior  in the limit $\tau \to +0$ and
 $\tau \to + \infty$  (where $\tau$ is the ``synchronous-type'' variable)
 in terms of inequalities on Kasner  parameters. We have also suggested
 examples of billiards related  to the hyperbolic Kac-Moody
 algebras  $E_{10}$, $AE_3$ and $A_{1,II}$.


 \begin{center}
 {\bf Acknowledgments}
 \end{center}

 This work was supported in part by the Russian Foundation for
 Basic Research grant  Nr. $07-02-13624-ofi_{ts}$.


\end{document}